\documentclass[comsoc]{IEEEtran}

\usepackage[draft]{todonotes}
\usepackage{pifont} 
\usepackage{lstautogobble}  
\usepackage{booktabs, makecell,threeparttable}

\definecolor{keywords}{RGB}{255,0,90}
\definecolor{comments}{RGB}{0,0,113}
\definecolor{red}{RGB}{160,0,0}
\definecolor{green}{RGB}{0,150,0}
\definecolor{airforceblue}{rgb}{0.36, 0.54, 0.66}
\definecolor{wildwatermelon}{rgb}{0.99, 0.42, 0.52}
\definecolor{viridian}{rgb}{0.25, 0.51, 0.43}
\definecolor{twilightlavender}{rgb}{0.54, 0.29, 0.42}
\definecolor{umber}{rgb}{0.39, 0.32, 0.28}
\definecolor{tropicalrainforest}{rgb}{0.0, 0.46, 0.37}
\definecolor{darkmagenta}{rgb}{0.55, 0.0, 0.55}

\definecolor{bluekeywords}{rgb}{0.13, 0.13, 1}
\definecolor{greencomments}{rgb}{0, 0.5, 0}
\definecolor{redstrings}{rgb}{0.9, 0, 0}
\definecolor{graynumbers}{rgb}{0.5, 0.5, 0.5}
\definecolor{alizarin}{rgb}{0.82, 0.1, 0.26}

\definecolor{figuregreen}{RGB}{141,190,133}
\definecolor{figureblue}{RGB}{141,188,235}

\definecolor{t1}{RGB}{215,25,28}
\definecolor{t2}{RGB}{253,174,97}
\definecolor{t3}{RGB}{255,255,191}
\definecolor{t4}{RGB}{171,221,164}
\definecolor{t5}{RGB}{43,131,186}

\lstdefinestyle{test}{ %
  basicstyle=\footnotesize,        
  breakatwhitespace=false,         
  breaklines=true,                 
  captionpos=b,                    
  commentstyle=\color{tropicalrainforest}\bfseries,
  deletekeywords={...},            
  escapeinside={\%*}{*)},          
  extendedchars=true,              
  frame=l,	                   	   
  keepspaces=true,                 
  keywordstyle=\color{red}\bfseries,  
  language=c,                 
  otherkeywords={*,...},           
  numbers=left,                    
  numbersep=5pt,                   
  rulecolor=\color{black},         
  showspaces=false,                
  showstringspaces=false,          
  showtabs=false,                  
  stepnumber=1,                    
  stringstyle=\color{darkmagenta}, 
  tabsize=4,	                   
  xleftmargin=\parindent,
  columns=flexible                    
}

\lstdefinestyle{py}{ %
   aboveskip=1ex,
   belowskip=1ex,
  basicstyle=\footnotesize,        
  breakatwhitespace=false,         
  breaklines=true,                 
  captionpos=b,                    
  columns=fullflexible,
  commentstyle=\color{tropicalrainforest}\bfseries,
  deletekeywords={...},            
  escapeinside={\%*}{*)},          
  extendedchars=true,              
  frame=l,	                   	   
  keepspaces=true,                 
  keywordstyle=\color{red}\bfseries,  
  language=python,                 
  otherkeywords={*,...},           
  numbers=left,                    
  numbersep=5pt,                   
  rulecolor=\color{black},         
  showspaces=false,                
  showstringspaces=false,          
  showtabs=false,                  
  stepnumber=1,                    
  stringstyle=\color{darkmagenta}, 
  tabsize=4,	                   
  xleftmargin=\parindent,
  columns=flexible                    
}

\lstdefinestyle{new}{
    backgroundcolor=\color{yellow!8},%
    aboveskip=2ex,
    belowskip=2ex,
    autogobble,
    columns=fullflexible,
    language=c,
    showspaces=false,
    showtabs=false,
    breaklines=true,
    showstringspaces=false,
    breakatwhitespace=true,
    escapeinside={(*@}{@*)},
    commentstyle=\color{greencomments}\bfseries,
    keywordstyle=\color{bluekeywords}\bfseries,
    stringstyle=\color{darkmagenta},
    numbers=left,
    numberstyle=\color{graynumbers},
    basicstyle=\ttfamily\footnotesize,
    frame=3,
    framesep=12pt,
    xleftmargin=12pt,
    tabsize=4,
    captionpos=b,
    literate={0}{{\textcolor{orange}{0}}}{1}%
             {1}{{\textcolor{orange}{1}}}{1}%
             {2}{{\textcolor{orange}{2}}}{1}%
             {3}{{\textcolor{orange}{3}}}{1}%
             {4}{{\textcolor{orange}{4}}}{1}%
             {5}{{\textcolor{orange}{5}}}{1}%
             {6}{{\textcolor{orange}{6}}}{1}%
             {7}{{\textcolor{orange}{7}}}{1}%
             {8}{{\textcolor{orange}{8}}}{1}%
             {9}{{\textcolor{orange}{9}}}{1}%
             {.0}{{\textcolor{orange}{.0}}}{2}
             {.1}{{\textcolor{orange}{.1}}}{2}
             {.2}{{\textcolor{orange}{.2}}}{2}%
             {.3}{{\textcolor{orange}{.3}}}{2}%
             {.4}{{\textcolor{orange}{.4}}}{2}%
             {.5}{{\textcolor{orange}{.5}}}{2}%
             {.6}{{\textcolor{orange}{.6}}}{2}%
             {.7}{{\textcolor{orange}{.7}}}{2}%
             {.8}{{\textcolor{orange}{.8}}}{2}%
             {.9}{{\textcolor{orange}{.9}}}{2}%
}

\lstdefinestyle{new-python}{
    backgroundcolor=\color{yellow!8},%
    aboveskip=2ex,
    belowskip=2ex,
    autogobble,
    columns=fullflexible,
    language=python,
    showspaces=false,
    showtabs=false,
    breaklines=true,
    showstringspaces=false,
    breakatwhitespace=true,
    escapeinside={(*@}{@*)},
    commentstyle=\color{greencomments}\bfseries,
    keywordstyle=\color{bluekeywords}\bfseries,
    stringstyle=\color{darkmagenta},
    morekeywords={Enum},
    numbers=left,
    numberstyle=\color{graynumbers},
    basicstyle=\ttfamily\footnotesize,
    frame=3,
    framesep=12pt,
    xleftmargin=12pt, 
    tabsize=4,
    captionpos=b,
    literate={0}{{\textcolor{orange}{0}}}{1}%
             {1}{{\textcolor{orange}{1}}}{1}%
             {2}{{\textcolor{orange}{2}}}{1}%
             {3}{{\textcolor{orange}{3}}}{1}%
             {4}{{\textcolor{orange}{4}}}{1}%
             {5}{{\textcolor{orange}{5}}}{1}%
             {6}{{\textcolor{orange}{6}}}{1}%
             {7}{{\textcolor{orange}{7}}}{1}%
             {8}{{\textcolor{orange}{8}}}{1}%
             {9}{{\textcolor{orange}{9}}}{1}%
             {.0}{{\textcolor{orange}{.0}}}{2}
             {.1}{{\textcolor{orange}{.1}}}{2}
             {.2}{{\textcolor{orange}{.2}}}{2}%
             {.3}{{\textcolor{orange}{.3}}}{2}%
             {.4}{{\textcolor{orange}{.4}}}{2}%
             {.5}{{\textcolor{orange}{.5}}}{2}%
             {.6}{{\textcolor{orange}{.6}}}{2}%
             {.7}{{\textcolor{orange}{.7}}}{2}%
             {.8}{{\textcolor{orange}{.8}}}{2}%
             {.9}{{\textcolor{orange}{.9}}}{2}%
}

\begin{document}
\bstctlcite{IEEEexample:BSTcontrol}

\title{
Transport Services: A Modern API for\\ an Adaptive Internet Transport Layer
}

\author{Michael Welzl, Safiqul Islam, Michael Gundersen, and Andreas Fischer

\thanks{\it Michael~Welzl and Safiqul~Islam are with the Department of Informatics, University of Oslo. E-mail: \{michawe, safiquli\}@ifi.uio.no.}%
\thanks{\it Michael~Gundersen is with Bekk. E-mail: michael.gundersen@bekk.no}%
\thanks{\it Andreas~Fischer is with the Fakult\"at Angewandte Informatik, Technische Hochschule Deggendorf. E-mail: andreas.fischer@th-deg.de.}%
}


\begin{minipage}{\textwidth}
  \vspace{5cm}

  \Large
  
  \emph{Michael Welzl, Safiqul Islam, Michael Gundersen, Andreas Fischer: "Transport Services: A Modern API for an Adaptive Internet Transport Layer", accepted for publication, IEEE Communications Magazine, April 2021.}

  \LARGE
  
  
  \vspace{50mm}
  Version accepted for publication in IEEE Communications Magazine.
  
  \vspace{10mm}
  \emph{\copyright 2021 IEEE. Personal use of this material is permitted. Permission from IEEE must be
    obtained for all other uses, in any current or future media, including
    reprinting/republishing this material for advertising or promotional purposes, creating new
    collective works, for resale or redistribution to servers or lists, or reuse of any copyrighted
    component of this work in other works.}
\end{minipage}
\clearpage

\maketitle

\begin{abstract}
Transport services (TAPS) is a working group of the Internet's standardization body, the Internet Engineering Task Force (IETF). TAPS defines a new recommended API for the Internet's transport layer. This API gives access to a wide variety of services from various protocols, and it is protocol-independent: the transport layer becomes adaptive, and applications are no longer statically bound to a particular protocol and/or network interface. We give an overview of the TAPS API, and we demonstrate its flexibility and ease of use with an example using a Python-based open-source implementation.
\end{abstract}



\section*{Introduction}

Previously, it was common to regard the Berkeley Software Distribution (BSD) Socket interface (also known as ``Berkeley sockets'') as the standard API
for the transport layer. Since the 1980's, it has been the best known way to access the two commonly
available transport protocols: TCP and UDP. Nowadays, however, more protocols and mechanisms
are available, offering a much richer set of services --- Multipath TCP (MPTCP) can
intelligently utilize multiple network paths~\cite{mptcp}; QUIC can multiplex independent data streams to avoid head-of-line blocking delay (among many other things)~\cite{quic};
Low Extra Delay Background Transport (LEDBAT) is a congestion control mechanism that enables ``background'' communication which
gets out of the way of ``foreground'' traffic~\cite{Welzl:2013:LEDBATsurvey}.

Today, these protocols and mechanisms are implemented and
used by industry giants: MPTCP is implemented in
iOS, and used by Apple in support of their applications  ``Siri'' and ``Apple Music''\footnote{MPTCP is now also supported by Network.Framework, which we will discuss later.}~\cite{WWDC2020};
QUIC is implemented and used in Google's Chrome browser~\cite{quic};
a variant of LEDBAT is implemented and used by Microsoft for Operating System updates, telemetry, and error reporting~\cite{ledbatplusplus}. Such in-house development of both the protocols and their use cases is very labour-intensive (and hence costly), leaving an important question unanswered: \emph{how can ``smaller'' users, such as small and medium-sized enterprises (SMEs), access these new services and benefit from them?}

\begin{figure}[t]
    \centering
    \includegraphics[width=\columnwidth]{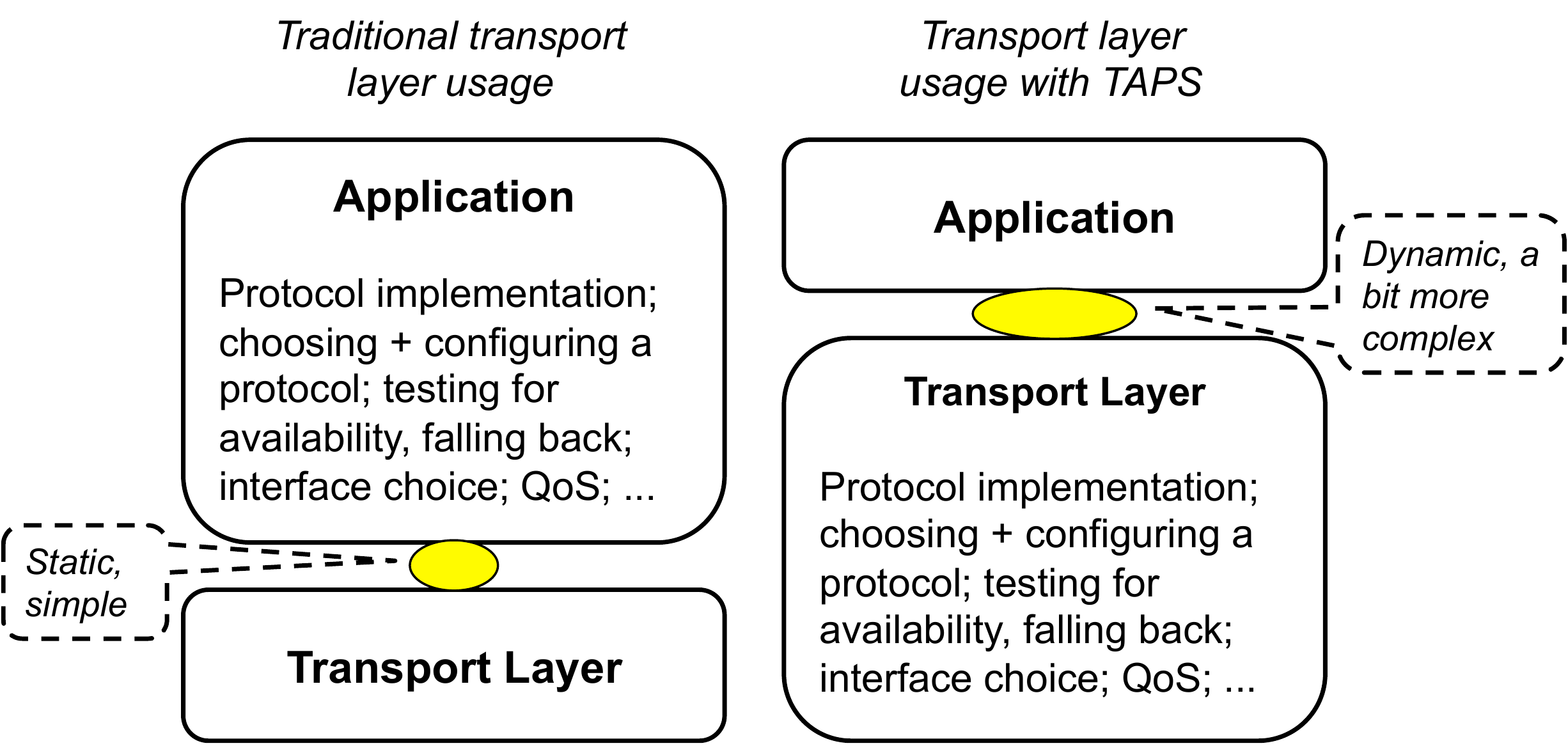}
  \caption{\small TAPS relocates the code to implement a protocol and/or choose and configure protocols, test for availability, etc., into the transport layer.}
  \vspace{-0.4cm}
  \label{fig:basicidea}
\end{figure}

Enter TAPS: The Transport Services working group (TAPS WG) of the Internet Engineering Task Force (IETF) intends to eliminate the static compile-time binding between applications and transport protocols, allowing each one of these elements to evolve independently --- and in doing so, it can put an end to the situation of growing unfairness between ``big'' and ``small'' developers. The basic idea, as shown in Fig.\,\ref{fig:basicidea}, is to move functionality out of the applications, into a common transport layer implementation (which resides in an Operating System or a library),
and to enable access to these functions via a new API. This 
has the potential to empower ``smaller players'' with a much richer set of services than ever before, but without the need to invest a huge amount of manpower into the development of a custom-made protocol.

By offering a modern API that follows an event-driven programming model, TAPS is also an effort to define a
new standardized interface to the transport layer for
programmers that is easy to use.
Compared to BSD sockets, which represent an old
fashioned, heavily C-influenced low-level programming style,
this should lower the entry barrier for network programmers seeking more services than common higher-level APIs such as an HTTP-based REST interface can offer.

We will now give an overview of the new Transport Services concept and how it affects the way network code is written; then, we will discuss available implementations of TAPS-conforming transport systems. Finally, we will use a Python example from the open-source implementation ``NEATPy'' to show the flexibility and ease of use of this new API.


\section*{Transport Services (TAPS) Overview}
\label{sec:taps}

As Fig.\,\ref{fig:basicidea} illustrates,
providing a flexible transport layer with interchangeable protocols and a run-time choice of network interfaces and protocol configurations requires a somewhat sophisticated machinery that dynamically and intelligently utilizes the available infrastructure. This machinery needs to take care of:
\begin{itemize}
    \item \emph{Protocol racing:} Peer and path support for protocols needs to be actively tested. Intelligent caching strategies should be used to limit such probing to save time and reduce server overhead.
    \item \emph{API-protocol mapping:} finding the matching functionality to support API calls can be a simple matter of calling function $X$ when request $Y$ is made, but it can also entail a more complicated use of underlying protocols.
    \item \emph{System policy management:} an application may express a wish to use a certain network interface --- yet, for example, smart phones commonly give the user system-wide control over the choice of the WiFi and cellular network interfaces. System controls are normally expected to overrule per-application preferences. Interface choice is only one  example of a system policy that may need to interact with an application preference; clearly, a richer API that offers applications a wealth of network mechanisms to choose and configure must meaningfully interact with the underlying system's policy manager.
\end{itemize}

This article focuses on the API. Thus, we refer to related work for further details on ``under the hood'' functionality. Reference~\cite{neat} gives an overview of the internals of the ``NEATPy'' implementation that we will discuss later, and general TAPS implementation guidance can be found in~\cite{ietf-taps-impl-07}.

\subsection*{Motivation}

We must first understand why there is a need for an API change at all. For example,
using the Stream Control Transmission Protocol (SCTP), it is possible to transparently map TCP connections between the same end hosts onto streams of a single association (SCTP's term for a connection)~\cite{flowmapping}.
This allows applications to benefit from a new protocol feature without changing the API (multi-streaming, which is available in SCTP and QUIC, reduces the signaling overhead and allows multiple data streams to jointly be controlled by a single congestion control instance).
There are, however, limits to the gains that can be obtained in such a way --- some protocol mechanisms \emph{must} be exposed in an API.

\subsubsection*{Head-of-line blocking example}

\begin{figure}[t]
    \centering
    \includegraphics[width=0.9\columnwidth]{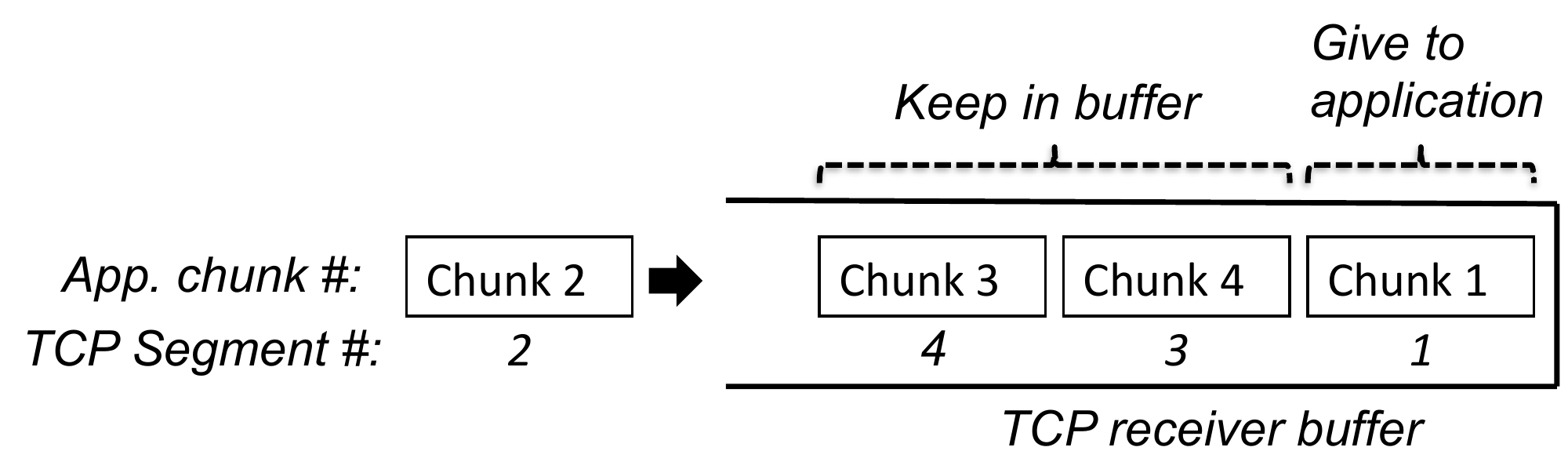}
  \caption{\small Application data chunks arriving at the TCP receiver in the wrong order. TCP segment 2, carrying chunk 2, arrives late, delaying the application's reception of chunks 3 and 4.}
  \vspace{-0.4cm}
  \label{fig:unordered}
\end{figure}

Consider an online multiplayer game that needs to reliably transfer position updates. The game may not care about the order of these updates, but they are latency-critical. Now, let us assume that this application uses TCP, and that every application data chunk fits inside exactly one TCP segment (this may be an unrealistic simplification for position updates in a game, which are typically very small, but the same arguments hold if a TCP segment contains multiple application data chunks). This situation is shown in Fig.\,\ref{fig:unordered}: here, four application data chunks are transmitted as TCP segments 1-4, and segment 2 arrives late (e.g., it was lost and retransmitted). Since TCP delivers data to the application as a consecutive byte stream, chunks 3 and 4 cannot be handed over; they have to wait in the TCP receiver buffer until segment 2 arrives.

Clearly, our game application could better be supported by a transport protocol that can hand over messages out of order --- but, most importantly, the protocol would \emph{need to be told about the size of messages, the requirement of reliable delivery, and the possibility to accept messages out-of-order}.
A drop-in TCP replacement below the BSD socket API could never hand over chunks 3 and 4: this would not be in line with the interface's expected behavior, and the application might fail. However, falling back to TCP (in case a different protocol is not available) would work: if an API allows out-of-order delivery, yet TCP is used below, then ordering will be ensured at the cost of efficiency. Ordered delivery is never incorrect, it may just be slower --- and, in line with the Internet's ``best effort'' service model, efficiency is not \emph{guaranteed}.

This example has shown us that a better transport layer \emph{must} offer some services beyond the well-known two: reliable byte stream (TCP) and unreliable datagram (UDP). These services must be reflected in all APIs in order to be usable: if, say, the socket API is extended to support this functionality, yet an application uses a communication library on top which only offers a consecutive byte stream, then, again, there is no way to benefit from unordered reliable message delivery. This means that upgrading the transport layer is not ``merely'' an API change --- it is a new way of \emph{thinking about} communication.

\subsection*{API Overview}
\label{sec:taps:keyapi}

Using BSD sockets requires working in a C-oriented low-level programming style of the 1980's (a socket has to be actively polled for incoming traffic, error codes are returned as integer values, etc.). This has contributed to a shift towards using other communication libraries or middle-ware systems that are built on top of BSD sockets. The services that such upper layers can expose are, however, inevitably limited by the underlying services (TCP and UDP). Hence, when changing the transport layer one should take the opportunity for a much-needed modernization of the interface.

Accordingly, the design of TAPS follows a more modern paradigm of network communication. It is fully asynchronous, event-driven, and easy to use: rather than distinguishing between a ``stream'' and ``datagram'' communication model, in TAPS, all data are transferred in the form of messages, and all communication is connection oriented. A ``connection'' is defined as ``shared state of two or more end systems that persists across messages that are transmitted between these end systems''; under the hood, a TAPS connection may be realized by UDP datagrams or TCP connections.

\subsubsection*{Control flow}
Communication begins with making decisions about the remote end to communicate with, specified in any way that is suitable (e.g., a DNS name, or ``any'', when listening), as well as transport properties and security parameters. Then, a ``preconnection'' is created. All of this should be done as early as possible, to give the transport system the necessary information to efficiently race protocols. Most transport properties to be used at this stage have a ``preference'' qualifier, with possible values \emph{require}, \emph{prefer}, \emph{ignore}, \emph{avoid} or \emph{prohibit}. \emph{Require} and \emph{prohibit} should be
used with care, as they limit the system's flexibility. Transport properties convey service requests, such as the use of a protocol that supports reliability, possibly configurable per message, and being able to use ``0-RTT'' session establishment (i.e., sending the first message without a preceding handshake).

Event handlers must be registered before connecting or listening. Similarly, if they are used, framers must be added to the preconnection at this time. Framers are functions that an application can define to translate data into application-layer messages; these functions will intercept all send() and receive() calls on the connection. In this way, an application can define its own message delineation (typically a protocol header --- e.g., the HTTP header, in case of an HTTP/TAPS implementation) that will allow the transport system to handle messages even when the underlying protocol is TCP.

Then, a connection is established by calling either the ``Initiate'' or ``InitiateWithSend'' primitive associated with the preconnection (or ``Listen'', in case of a server). The semantics of ``Initiate'' are slightly different from the traditional ``connect'' and ``accept'': calling ``Initiate'' is not guaranteed to invoke a ``ConnectionReceived'' event (the TAPS equivalent of 
``accept'') at the peer --- for example, in case of UDP, ``ConnectionReceived'' occurs when the first message arrives.

Data are always transferred as messages. Each message may have associated properties to express requirements such as a lifetime, reliability, ordering, etc. Connections also have properties that can be changed while they are active --- for example, a capacity profile, which can influence the value of the DiffServ CodePoint (DSCP) in the IP header.

On the sender side, it is possible to execute some level of control over the send buffer because a ``sent'' event is fired when a send action has completed (i.e., the message has been passed to the underlying protocol stack), and these ``sent'' events can be used to steer data transmission --- for example, allowing only one message to be buffered at a time by issuing one ``send'' per ``sent''. Applications that focus on traffic that is not latency critical may simply ignore these ``sent'' events. On the receiver side, it is necessary to avoid being overwhelmed by too many quickly firing ``received'' events. This is achieved via the ``receive'' call, which queues the reception of a message; the system guarantees a one-to-one mapping between ``receive'' calls and ``received'' events.

\begin{figure}[t]
    \centering
    \includegraphics[width=0.85\columnwidth]{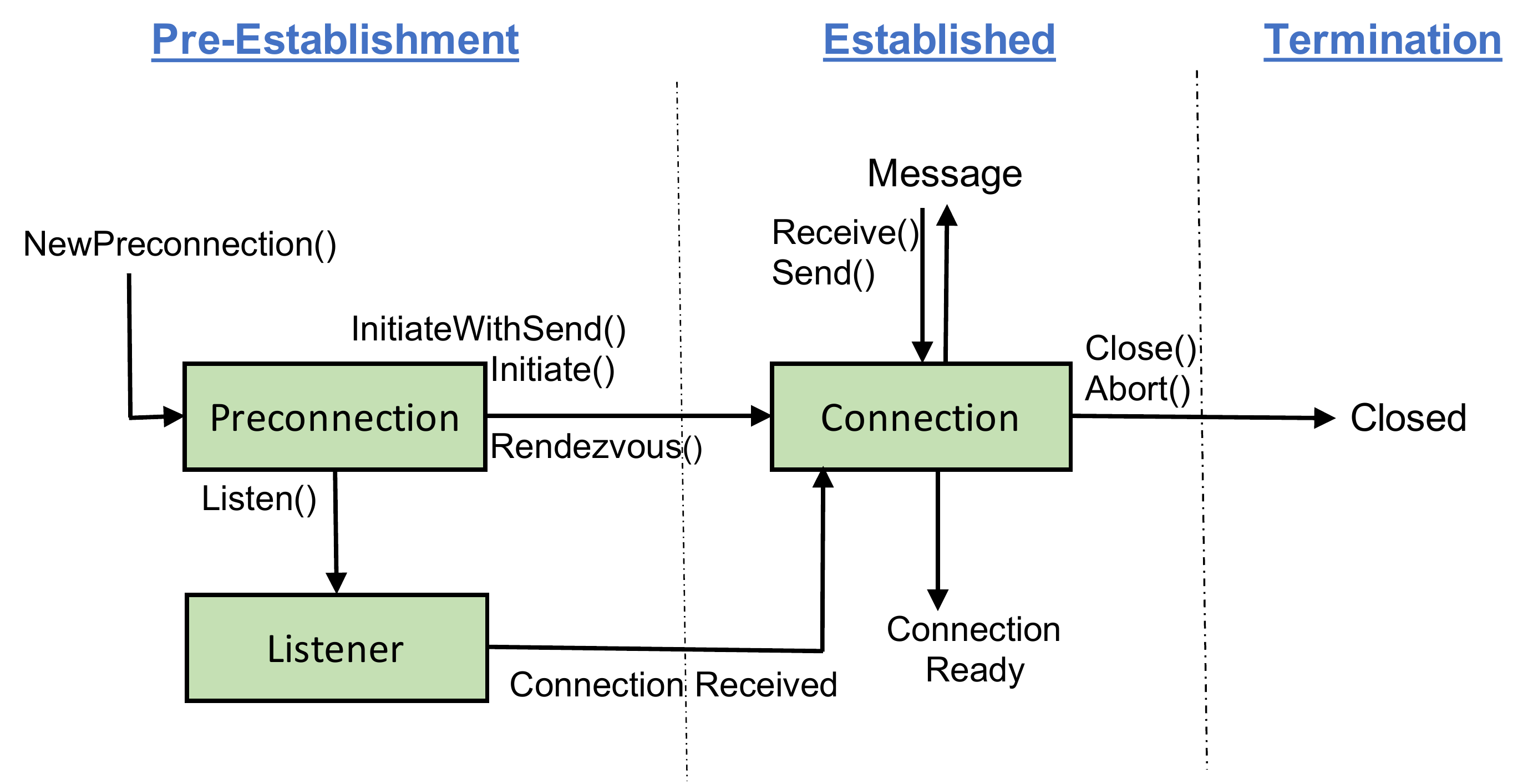}
  \caption{\small Lifetime of a connection provided by a TAPS transport system (redrawn from~\cite{ietf-taps-arch-08}).}
  \vspace{-0.4cm}
  \label{fig:controlflow}
\end{figure}
Closing a connection is also not guaranteed to invoke an event at the peer: in case of TCP, it will, but in case of UDP, it will not. Half-closed connections (as with TCP) are not supported because not all protocols support them (e.g., SCTP does not), and supporting half-closed connections would therefore prohibit the use of these protocols. Figure\,\ref{fig:controlflow} gives a high-level overview of the control flow (connection lifetime) that we have just described.

\subsubsection*{Cloning}
For new connections that are established to an already connected peer, it is recommended to use the ``Clone'' primitive with the ongoing connection. A successful ``Clone'' call will yield a new connection that is ``entangled'' with the existing one; changing a property on one of them will affect the other. This continues similarly for all connections of a group that is formed by calling 
``Clone'' multiple times. Using ``Clone'' allows the transport system to represent a new connection as a new stream of an existing underlying transport connection or association if the protocol supports this.

A TAPS transport system can fall back to TCP in case no newer protocol is supported by the peer and the path. This enables one-sided deployment of new protocols. ``Clone'' may therefore yield a new TCP connection; to avoid surprises, the system will then ensure that changing a connection property affects all the connections in a group.
Specifying a capacity profile, or allowing \emph{un}ordered or \emph{un}reliable message delivery may not have an effect. None of these fall-backs endanger correctness --- using TCP instead of a desired better protocol merely sacrifices performance.


\section*{Implementations}
\label{sec:implementations}

\begin{table*}[!ht]
    \centering
    \footnotesize
    \renewcommand{\arraystretch}{1.5}
    \begin{tabular}{cccc}
        \toprule[1pt]
        \textbf{Support for...} & \textbf{PyTAPS} & \textbf{Network.Framework} & \textbf{NEATPy} \\
        \midrule[1pt]
        TCP/IP & \ding{51} & \ding{51} & \ding{51}\\ 
        UDP/IP & \ding{51} & \ding{51}  &  \ding{51}\\ 
        SCTP/IP &  \ding{55}&  \ding{55}& \ding{51}\\ 
        STCP/UDP/IP &\ding{55}  &\ding{55}  & \ding{51}\\ 
        TLS/TCP/IP & \ding{51} & \ding{51} & \ding{51} \\ 
        DTLS/UDP/IP & \ding{55} & \ding{51} & \ding{51} \\ 
        DTLS/SCTP/IP & \ding{55} & \ding{55} & \ding{51}\\ 
        DTLS/STCP/UDP/IP &\ding{55}  &\ding{55}  & \ding{51}\\ 
        MPTCP & \ding{55} & \ding{51} & \ding{51} \\ 
        Protocol selection by selection properties  & \ding{51} & \ding{55} & \ding{51}\\ 
        Transport protocol racing & \ding{51} & \ding{55} & \ding{51}\\ 
        Message framers & \ding{51} &  \ding{51} & \ding{51}\\ 
        Message context / Properties & \ding{55} & \ding{51} & \ding{51}\\ 
        Cloning  & \ding{55} & \ding{55} & \ding{51} \\ 
        Rendezvous & \ding{55} & \ding{51} & \ding{55}\\ 
        Connection properties & \ding{55} &\ding{51}  & \ding{51}\\ 
        \bottomrule[1pt]
    \end{tabular}
    \caption{\small The three known TAPS implementations: supported protocol stacks and key TAPS features.}
    \label{tab:impl}
\end{table*}

There are currently three known implementations of a transport system conforming to the IETF TAPS specification: 
PyTAPS \cite{pytaps}, Network.Framework \cite{networkframework}, and NEATPy \cite{neatpy}. Table\,\ref{tab:impl} shows a comparison of the three implementations and their respective protocol and feature support.

PyTAPS is a prototype implementation of a transport system, using the specification of the abstract interface by the IETF TAPS working group. PyTAPS supports TCP, UDP, and the use of TLS over TCP. 
It is written in Python and uses asyncio, a Python Standard Library for writing concurrent code. It presents an asynchronous
transport system with an event loop that operates on tasks. Concurrent execution is based on Python co-operative routines (coroutines). A coroutine  is an asynchronous block of code with the ability to ``yield'', that is, pause its execution and give control to the event loop scheduler at any time during its execution, and maintain its internal state. PyTAPS uses co-routines to define all API and callback functions.

Network.framework is Apple's reference implementation of a TAPS system. This implementation is available in both Objective-C and Swift, and it is used to transport application data across Apple's many platforms. However, Apple's implementation does not currently specify abstract requirements needed for the transportation of data, which could be used for the selection of a protocol that satisfies certain requirements. Instead, the user can indicate preferences 
tied to a specific protocol. For example, UDP is modeled as a class called \emph{NWProtocolUDP}, which has the option \emph{preferNoChecksum}. 
Naturally, being Apple's common network interface in production use, it offers many services beyond a common TAPS API. Examples include the possibility for developers to get highly detailed connection metrics and the ability to create connections using WebSockets.

NEATPy presents a Python-based TAPS API, realized with the help of language bindings, utilizing the protocol machinery of its underlying core system ``NEAT''.
NEAT (A New, Evolutive API and Transport-Layer Architecture for the Internet), which is described in detail in~\cite{neat},
was the first open-source implementation of a TAPS system, written in C. It was an output of the European research project with the same name. Development work on NEAT finished with the project's end, in 2018; in contrast, NEATPy's development persisted until mid-2020, bringing NEAT's core functionality in line with an up-to-date TAPS system. The NEAT API differs from the most recent TAPS specification in several ways. For example, while NEAT already allowed to specify selection and connection properties, it did not offer message properties---instead, some per-message functions were available as parameters of the send() call. Instead of the five preference levels of TAPS, NEAT only supported qualifying properties as ``required'' or ``desired''. Also, framers were not supported in NEAT---it was entirely up to the application to parse messages from an incoming block of data.

NEATPy can benefit from NEAT's policy component in the NEAT user module, which maps properties to policies. These policies do not only enable protocol racing between the candidate protocols but also provide a fallback mechanism in case a selection of a protocol fails.
NEATPy has more features and supports more protocol stacks (including SCTP with and without UDP encapsulation, and MPTCP; the latter requires installing the reference MPTCP implementation from~\cite{mptcp-impl}) than PyTAPS, but this comes at the cost of more overhead and slower overall execution.
NEATPy can run on various operating systems, and it will make use of different capabilities depending on what the underlying operating system offers. Our tests used NEATPy on Linux and FreeBSD.

For developers, the choice between the three different implementations should be relatively easy: Network.Framework is an obvious choice for Apple systems, where it is tied to the programming languages Objective-C and Swift. Else, if the intention is to use and possibly extend a lightweight system, PyTAPS should be preferred to NEATPy. The latter seems a good choice for the more experimentally minded, giving access to a wealth of features via a somewhat heavyweight underlying library. To date, SCTP is only supported by NEATPy, and QUIC is not presently supported by any of the three TAPS implementations, but it could be added.


\section*{TAPS in Action}
\label{sec:tapsinaction}

To show how a TAPS system operates, we present a code example of NEATPy and discuss the performance achieved in a local test using an emulated network environment.

\subsection*{Code: A Client-Server Example}

\begin{figure*}

\begin{minipage}[b]{\textwidth}
\begin{lstlisting}[label=fig:code,caption={\small A TAPS client and server example using a framer and two entangled connections in NEATPy, available from~\cite{neatpy}.},style=new-python]
#### SERVER ####

def simple_receive_handler(connection, message, context, is_end, error):
    data = message.data.decode()
    print(f"Got message with length {len(message.data)}: {data}")
    connection.send(data.encode("utf-8"), None)

def new_connection_received(connection: Connection):
    connection.receive(simple_receive_handler) 

if __name__ == "__main__":
    local_specifier = LocalEndpoint()
    local_specifier.with_port(5000)
    tp = TransportProperties()

    preconnection = Preconnection(local_endpoint=local_specifier, transport_properties=tp)
    new_listener = preconnection.listen()
    new_listener.HANDLE_CONNECTION_RECEIVED = new_connection_received

    preconnection.start()


#### CLIENT ####

class FiveBytesFramer(Framer):
    def start(self, connection): pass
    def stop(self, connection): pass

    def new_sent_message(self, connection, message_data, message_context, sent_handler, is_end_of_message):
        connection.message_framer.send(connection, 'HEADER'.encode("utf-8") + message_data,
                                       message_context, sent_handler, is_end_of_message)

    def handle_received_data(self, connection):
        header, context, is_end = connection.message_framer.parse(connection, 6, 6)
        connection.message_framer.advance_receive_cursor(connection, 6)
        connection.message_framer.deliver_and_advance_receive_cursor(connection, context, 5, True)

def clone_error_handler(con:Connection):
    print("Clone failed!")

def receive_handler(con:Connection, msg, context, end, error):
    print(f"Got message with length {len(msg.data)}: {msg.data.decode()}")

def ready_handler1(connection: Connection):
    connection.receive(receive_handler)   
    connection.send(("FIVE!").encode("utf-8"), None)
    connection2 = connection.clone(clone_error_handler)
    connection2.HANDLE_STATE_READY = ready_handler2

def ready_handler2(connection: Connection):
    connection.receive(receive_handler)    
    connection.send(("HelloWorld").encode("utf-8"), None)

if __name__ == "__main__":
    ep = RemoteEndpoint()
    ep.with_address("127.0.0.1")
    ep.with_port(5000)

    tp = TransportProperties()
    tp.require(SelectionProperties.RELIABILITY)
    tp.prohibit(SelectionProperties.PRESERVE_MSG_BOUNDARIES)   
 
    preconnection = Preconnection(remote_endpoint=ep, transport_properties=tp)
    preconnection.add_framer(FiveBytesFramer())
    connection1 = preconnection.initiate()
    connection1.HANDLE_STATE_READY = ready_handler1

    preconnection.start()
\end{lstlisting}
\end{minipage}
\end{figure*}

Listing\,\ref{fig:code} shows a TAPS server and client, implemented using NEATPy. The server is written to be as simple as possible, while we use the client to highlight a little bit more of the typical TAPS functionality. This code is runnable and complete except for some \verb'import' statements at the top, which are omitted for brevity.

The TAPS server listens to incoming connections and simply prints and returns any messages that it gets. A preconnection object is created, and two arguments are passed to it: a local endpoint (this specifies a port number where the server will listen) and a transport properties object (this sets a preference level for a couple of selection properties). Since no properties are configured in the transport properties object, this server will listen on all available protocols that support reliability (enabling reliability is a default property choice, as specified in~\cite{ietf-taps-interface-09}).

Then, we call listen() to accept any incoming connections from remote endpoints. The server uses two event handlers. The first event handler, ``new\_connection\_received'', is registered with the member \verb'HANDLE_CONNECTION_RECEIVED' of the listener class whenever a new connection is established, and the second event handler is registered inside the the first event handler when queuing a receive event. The second event handler receives the message, converts its bytes to text, prints the text to the screen and sends the data back.
Having configured the preconnection, registered the event handlers, and called ``listen'', we call the preconnection's start() method in order to start the transport system.

The client also creates a preconnection object, to which it passes a remote endpoint object (specifying the remote address and port) and transport properties. In this case, not only do we ask for reliable data transfer, but we prohibit the preservation of message boundaries, which practically enforces TCP---indeed, without this requirement, NEATPy communicated via SCTP in our test, and adding ``tp.ignore(SelectionProperties.PRESERVE\_ORDER)'' would give us the behavior that we discussed earlier (Fig.\,\ref{fig:unordered}).

Prohibiting message boundary preservation may be a strange request to make, but it allows us to test if a message framer works correctly even when the underlying transport protocol treats all data as a byte tream. To see this, we add an object of our \verb'FiveBytesFramer' class to the preconnection. This framer adds a textual header containing the word \verb'HEADER' to all messages (in the method \verb'new_sent_message'), which are supposed to contain only five bytes of data. Upon receiving (\verb'handle_received_data'), this header is removed, and the five bytes are handed over to the data reception handler.
The framer inherits from the abstract Framer class, which requires defining the ``start'' and ``stop'' methods; these allow to implement initialization and finalization activities, before/after any data are written or read. We leave them empty as we do not need such functionality in our example.

Back in the main function, we register the first event handler with \verb'HANDLE_STATE_READY' when a connection is established, and we call the \verb'start()' function to start the transport system.  This invokes \verb'ready_handler1' as soon as the connection is ready to accept data. There, one receive event is queued, a message containing the data \verb'FIVE!' is sent, and a new connection is created via \verb'clone'. Since we use TCP, this just produces another TCP connection, but with SCTP (in FreeBSD only, as support for this type of connection-stream mapping has not been implemented for Linux in the NEAT core), the new connection (\verb'connection2') would be a new stream of the already existing SCTP association to the server. The new connection's handler for the ready event also queues a single receive event and transmits a message, this time containing more than five bytes of data: \verb'HelloWorld'. Both connections use the same receive handler, which only prints out the received data together with its length.

Running this code produces the following output on the server side:

\noindent
\texttt{Got message with length 11: HEADERFIVE!\\
Got message with length 16: HEADERHelloWorld}

This output contains the header because the server does not implement a framer and simply prints out the raw message in full. On the client side, the output looks as follows:

\noindent
\texttt{Got message with length 5: FIVE!\\
Got message with length 5: Hello}

As we can see, the framer has removed the header upon reception, and the second message was correctly truncated to a length of five bytes.

\begin{figure}[t]
    \centering
    \includegraphics[width=0.7\columnwidth]{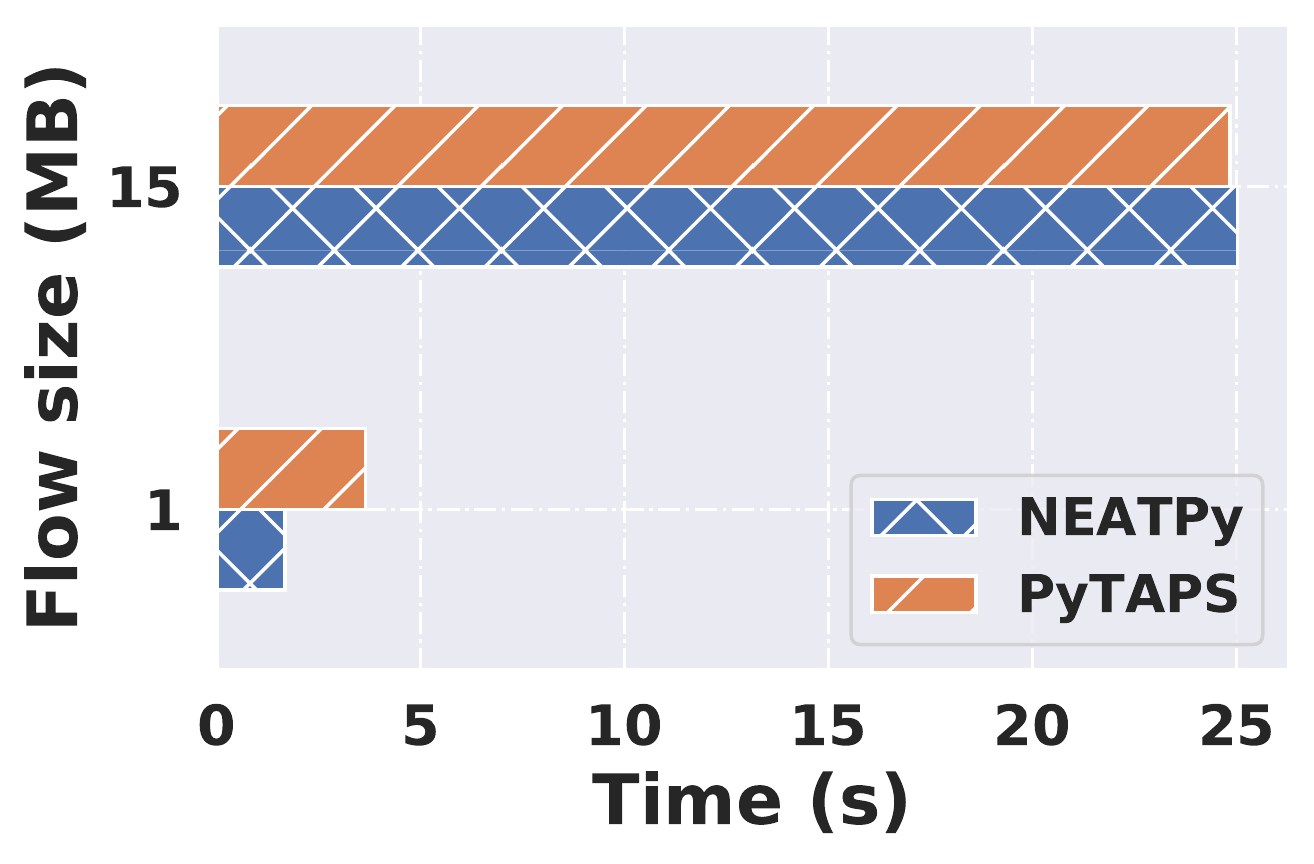}
    \caption{\small Flow completion time of a long (15\,MB) and short (1\,MB) flow with NEATPy and PyTAPS: the transfer time of a short flow with NEATPy, joining after 10 seconds, is significantly reduced because it benefits from the SCTP association's large congestion window.}
  \label{fig:multistreaming}
  \vspace{-0.4cm}
\end{figure}

\subsection*{Performance}
To demonstrate the benefit of protocol independent, portable code, we ran two simple experiments between a client and a server
on a single physical host,
using two instances of ``VirtualBox'' with FreeBSD OSes for a sender and a receiver, respectively. The two VirtualBox instances were logically interconnected on our Mac OS host system, and we set a maximum rate of 5\,Mbit/s and introduced a propagation delay of 30\,ms using dummynet/ipfw. We opted for FreeBSD because the second experiment uses multistreaming, which for NEATPy is only available with FreeBSD.

The first experiment is a simple ``hello world'' style test, where we merely transferred a single 12-byte message with PyTAPS and NEATPy, and found PyTAPS to be faster: the transfer took 0.202 seconds with NEATPy and 0.173 seconds with PyTAPS (this is the average duration of 10 tests, with a standard deviation of 4 percent).
This is not surprising: PyTAPS is an altogether much more lightweight implementation, and the reduced overhead plays out positively here.

The next experiment aims to show the benefit of a mechanism in a protocol that is available in NEATPy but not in PyTAPS: SCTP's multi-streaming. We used two connections, a long file transfer that is joined by a short file transfer after ten seconds, exploiting `clone' in case of NEATPy.

Figure\,\ref{fig:multistreaming} shows the result: multi-streaming yields a significant improvement in the short flow's completion time (FCT) because, being just a new stream of an ongoing SCTP association, it can immediately take advantage of the association's already-grown congestion window. The FCT of the short flow with NEATPy is reduced by 54 percent in comparison with the short flow with PyTAPS, where the two TAPS connections become two TCP connections, without support for multistreaming. We repeated this test ten  times with one long flow (15\,MB) starting at t=0\,s, and one short flow (1\,MB) starting at t=10\,s, and show the average FCT. The standard deviation was between 0,49 percent and 1.53 percent.

PyTAPS and NEATPy expose a very similar API (not 100 percent equal because they each have their own language-specific ways to implement the abstract interface specified in~\cite{ietf-taps-interface-09}). Thus, the code used in these two tests was essentially the same, with only minor syntactical changes. This means that (almost) the same program ran faster on the lighter-weight implementation when we did not utilize the protocol feature ``multi-streaming'', and it ran faster on the heavier implementation when we did use that feature. This is the flexibility that TAPS aims to attain: code can be portable, yet it can benefit from underlying protocol features that go beyond plain TCP and UDP.


\section*{Conclusion}
\label{sec:conclusion}

This article presented and discussed TAPS as a modern and flexible transport layer replacement for the legacy BSD Socket API.
At the time of writing, the Transport Services Working Group is close to finishing its three core documents: the architectural overview~\cite{ietf-taps-arch-08}, API~\cite{ietf-taps-interface-09} and implementation guidance~\cite{ietf-taps-impl-07}.
We discussed three implementations of this novel API and demonstrated its flexibility with code samples employing
NEATpy. This flexibility allows experimenters to easily switch between
implementations with only minor modifications to the code, while being able to exploit features of novel transport protocols that go well beyond TCP's reliable byte stream on one hand, and UDP's unreliable datagram transmission on the other.

TAPS implementations greatly facilitate the comparison of different transport protocols. Support for new protocols such as QUIC, or novel configurable extensions to existing protocols could be added to the modular open-source code of NEAT---and with it, NEATPy---relatively easily. This should make it an attractive tool for the research community.

\section*{Acknowledgments}

This work has been supported by the Research Council of Norway under its ``Toppforsk'' programme through the ``OCARINA'' project (grant agreement no. 250684).

\bibliographystyle{IEEEtran}




\begin{IEEEbiography}[{\includegraphics[width=1.05in,height=1.5in,clip,keepaspectratio]{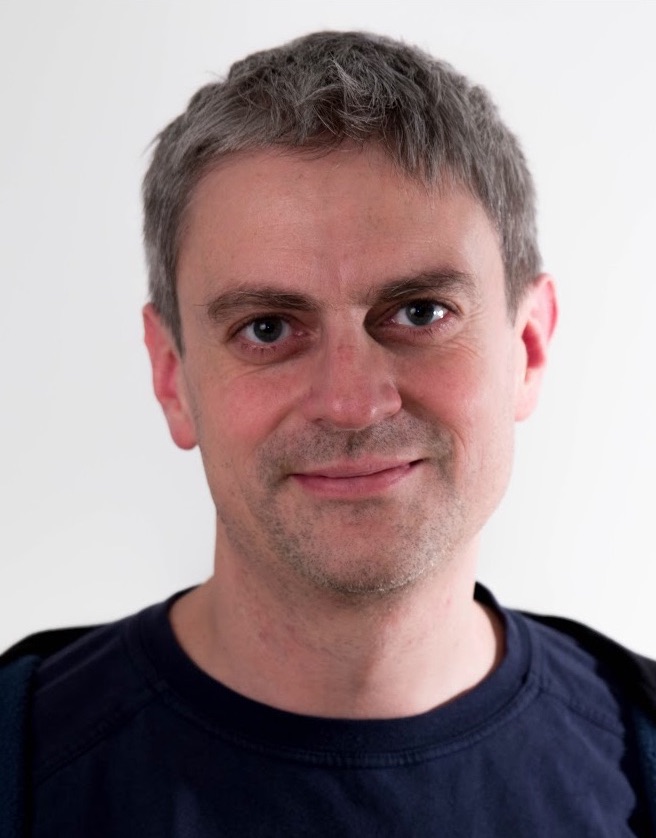}}]{\textsc{Michael Welzl}}
(michawe@ifi.uio.no) is a full professor at the University of Oslo,
Norway, since 2009. He received a Ph.D. and habilitation
from the University of Darmstadt / Germany in 2002 and 2007,
respectively. His main research focus is the
transport layer; he is active in the IRTF, where he chaired the Internet Congestion Control Research Group (ICCRG) for 11 years, and the IETF, where he led the initiative to form the TAPS Working Group.
\end{IEEEbiography}

\vspace{-20pt}
\begin{IEEEbiography}[{\includegraphics[width=1.05in,height=1.5in,clip,keepaspectratio]{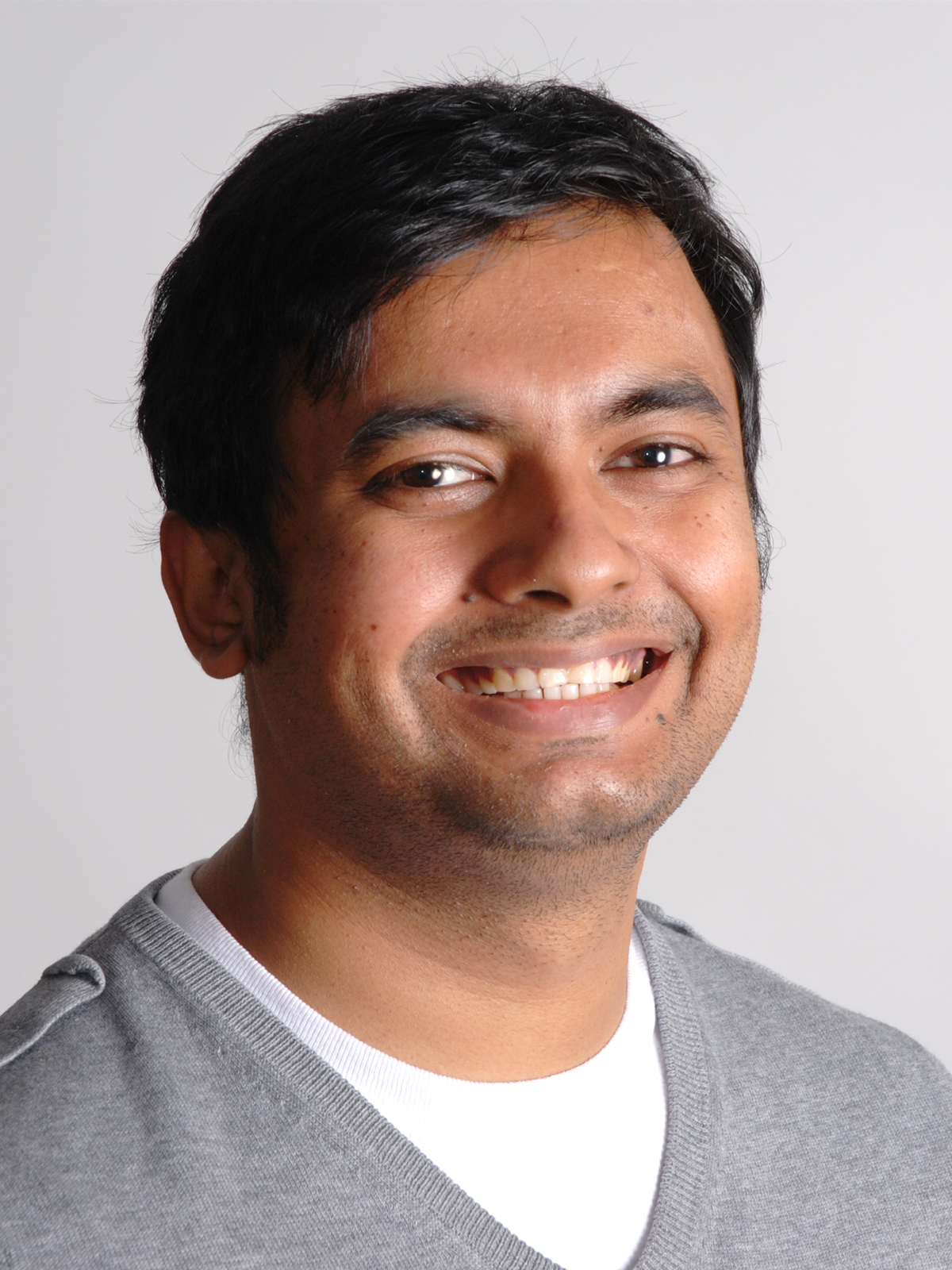}}]{\textsc{Safiqul Islam}}
(safiquli@ifi.uio.no) received a Ph.D. in Computer Science from the University of Oslo, Norway. Currently, he is a Postdoctoral Fellow at the Department of Informatics, University of Oslo. His research interests include performance analysis, evaluation, and optimization of transport layer protocols. He is active in the IETF and IRTF where he has contributed to several IETF/IRTF Working Groups.
\end{IEEEbiography}

\vspace{-20pt}
\begin{IEEEbiography}[{\includegraphics[width=1.05in,height=1.5in,clip,keepaspectratio]{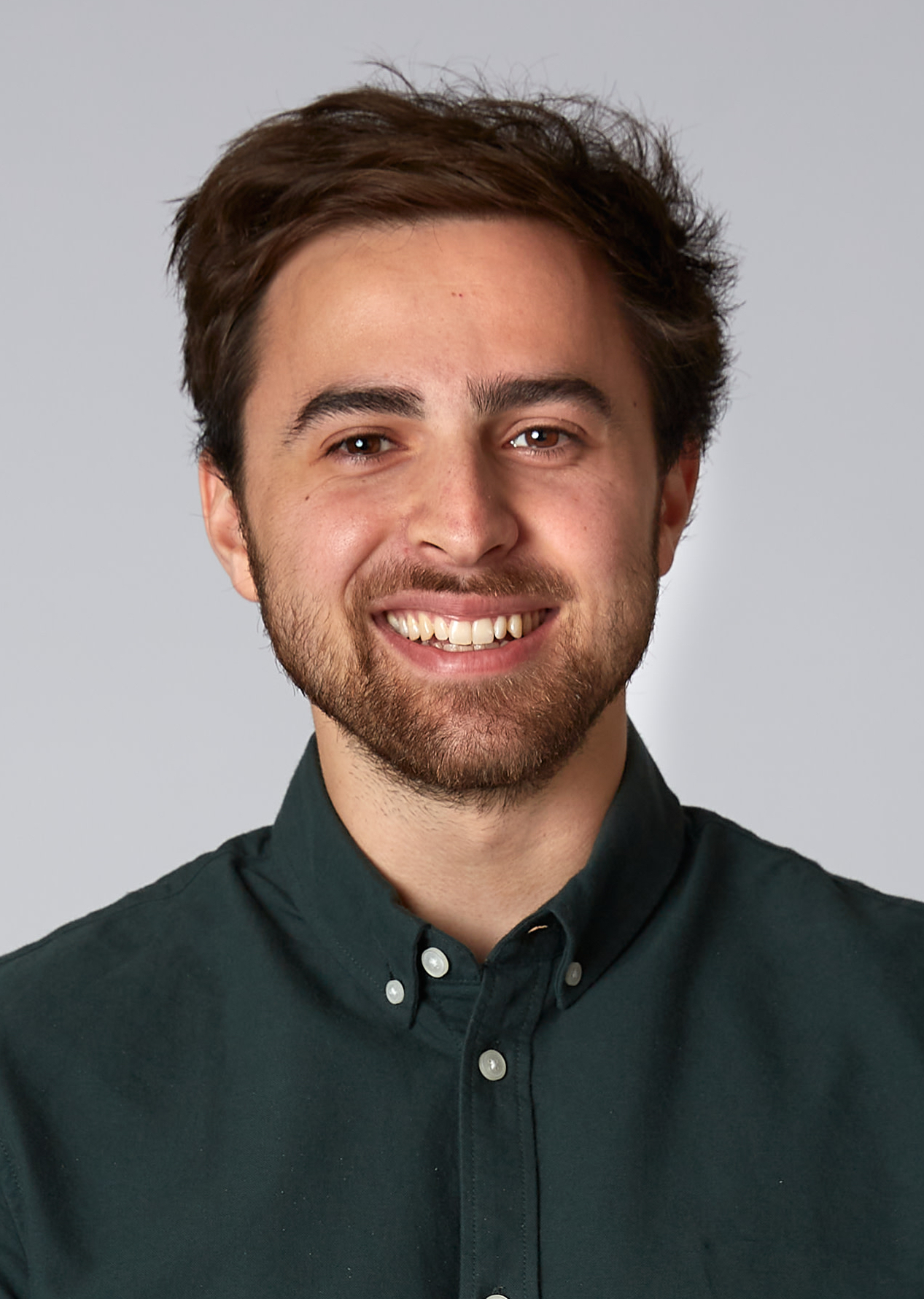}}]{\textsc{Michael Gundersen}}
(michael.gundersen@bekk.no) received an M.Sc. in Computer Science from the University of Oslo, Norway. He works as a developer at Bekk, a consultancy in Oslo specializing in technology, design and management. Among his main interests are API design and IoT. 
\end{IEEEbiography}

\vspace{-20pt}
\begin{IEEEbiography}[{\includegraphics[width=1.05in,height=1.5in,clip,keepaspectratio]{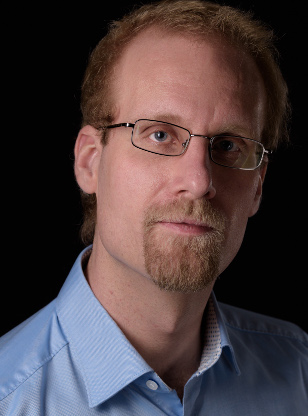}}]{\textsc{Andreas Fischer}}
(andreas.fischer@th-deg.de) is a Professor at Deggendorf Institute of Technology. He received a PhD in 2017
at University of Passau.  After a post-doc position in Karlstad, Sweden he was
appointed as professor for Computer Science at DIT. He is interested
in the development of intelligent and autonomous networks and has conducted
extensive research on network resilience, network virtualization and
software-defined networks. He has been active in the Future Internet community
for a long time.
\end{IEEEbiography}

\end{document}